\documentclass[letter]{aa} 
\usepackage{mhchem}
\usepackage{graphicx}
\usepackage{txfonts}
\usepackage[]{hyperref}
\hypersetup{
colorlinks=true,
linkcolor=blue,
urlcolor=cyan,
pdftitle={draft C2H5NCO},
citecolor=blue,
}
\usepackage[utf8]{inputenc}
\DeclareUnicodeCharacter{2212}{\circ}
\usepackage{nicefrac,xfrac}
\begin{document}


  \title{First detection of C$_2$H$_5$NCO in the ISM and search of other isocyanates towards the G+0.693-0.027 molecular cloud}
  \titlerunning{First detection of C$_2$H$_5$NCO in the ISM and search of other isocyanates towards G+0.693-0.027}

   \author{L.~F.~Rodr\'{\i}guez-Almeida
          \inst{1}
          \and
          V.~M.~Rivilla\inst{1,2}
          \and
          I.~Jim\'enez-Serra\inst{1}
          \and
          M.~Melosso\inst{3}
          \and
          L.~Colzi\inst{1,2}
          \and
          S.~Zeng\inst{4}
          \and
          B.~Tercero\inst{5}
          \and
          P.~de Vicente\inst{5}
          \and
          S.~Mart\'{\i}n\inst{6,7}
          \and
          M.~A.~Requena-Torres\inst{8,9}
          \and
          F.~Rico-Villas\inst{1}
          \and
          J.~Mart\'{\i}n-Pintado\inst{1}
          }

   \institute{Centro de Astrobiolog\'{\i}a (CSIC-INTA),
              Ctra Ajalvir km 4, 28850, Torrej\'{o}n de Ardoz, Madrid, Spain\\
              \email{lrodriguez@cab.inta-csic.es}
         \and
             INAF-Osservatorio Astrofisico di Arcetri,
             Largo Enrico Fermi 5, 50125, Florence, Italy
        \and
            Dipartimento di Chimica "Giacomo Ciamician", Universit\'a di Bologna, via F.~Selmi~2,40126, Bologna, Italy
        \and
          Star and Planet Formation Laboratory, Cluster for Pioneering Research, RIKEN,2-1 Hirosawa, Wako, Saitama, 351-0198, Japan
        \and
          Observatorio de Yebes (IGN), Cerro de la Palera s/n, 19141, Guadalajara, Spain
        \and 
            Eureopean Southern Observatory, Alonso de C\'{o}rdova 3107, Vitacura 763 0355, Santiago, Chile
        \and
            Joint ALMA Observatory, Alonso de C\'{o}rdova 3107, Vitacura 763 0355, Santiago, Chile
        \and
        University of Maryland, College Park, ND 20742-2421, USA
        \and
        Department of Physics, Astronomy and Geosciences, Towson University, MD 21252, USA     
             }

   \date{Received XX, YY; accepted ZZ, YY}

 
  \abstract
   {Little is known about the chemistry of isocyanates (compounds with the functional group R-N=C=O) in the interstellar medium, as only four of them have been detected so far: isocyanate radical (NCO), isocyanic acid (HNCO), N-protonated isocyanic acid (H$_2$NCO$^+$) and methyl isocyanate (CH$_3$NCO). The molecular cloud G+0.693-0.027, located in the Galactic Centre, represents an excellent candidate to search for new isocyanates since it exhibits high abundances of the simplest ones, HNCO and CH$_3$NCO.}
  {After CH$_3$NCO, the next complex isocyanates are ethyl isocyanate (C$_2$H$_5$NCO) and vinyl isocyanate (C$_2$H$_3$NCO). Their detection in the ISM would enhance our understanding of the formation of these compounds in space.}
   {We have searched for C$_2$H$_5$NCO, H$_2$NCO$^+$, C$_2$H$_3$NCO and cyanogen isocyanate (NCNCO) in a sensitive unbiased spectral survey carried out in the 2~mm and 7~mm radio windows using IRAM 30m and Yebes 40m radio telescopes, respectively.}
   {We have detected C$_2$H$_5$NCO and H$_2$NCO$^+$ towards G+0.693-0.027 (the former for the first time in the interstellar medium) with  molecular abundances of (4.7$-$7.3)$\times$10$^{-11}$ and (1.0$-$1.5)$\times$10$^{-11}$, respectively. A ratio CH$_3$NCO~/~C$_2$H$_5$NCO~=~8$\pm$1 is obtained; therefore the relative abundance determined for HNCO:CH$_3$NCO:C$_2$H$_5$NCO is 1:1/55:1/447, which implies a decrease by more than one order of magnitude going progressively from HNCO to CH$_3$NCO and to C$_2$H$_5$NCO. This is similar to what has been found for e.g. alcohols and thiols and suggests that C$_2$H$_5$NCO is likely formed on the surface of dust grains. In addition, we have obtained column density ratios of
   HNCO~/~NCO~>~269, HNCO~/~H$_2$NCO$^+$~$\sim$~2100 and C$_2$H$_3$NCO~/~C$_2$H$_5$NCO~<~4. A comparison of the Methyl~/~Ethyl ratios for isocyanates (-NCO), alcohols (-OH), formiates (HCOO-), nitriles (-CN) and thiols (-SH) is performed and shows that ethyl-derivatives may be formed more efficiently for the N-bearing molecules than for the O- and S-bearing molecules.
   }
   {}

   \keywords{Astrochemistry --
                ISM: molecules -- line: identification
               }

   \maketitle
%

\begin{figure*}[h]
   \centering
    \includegraphics[width=15cm]{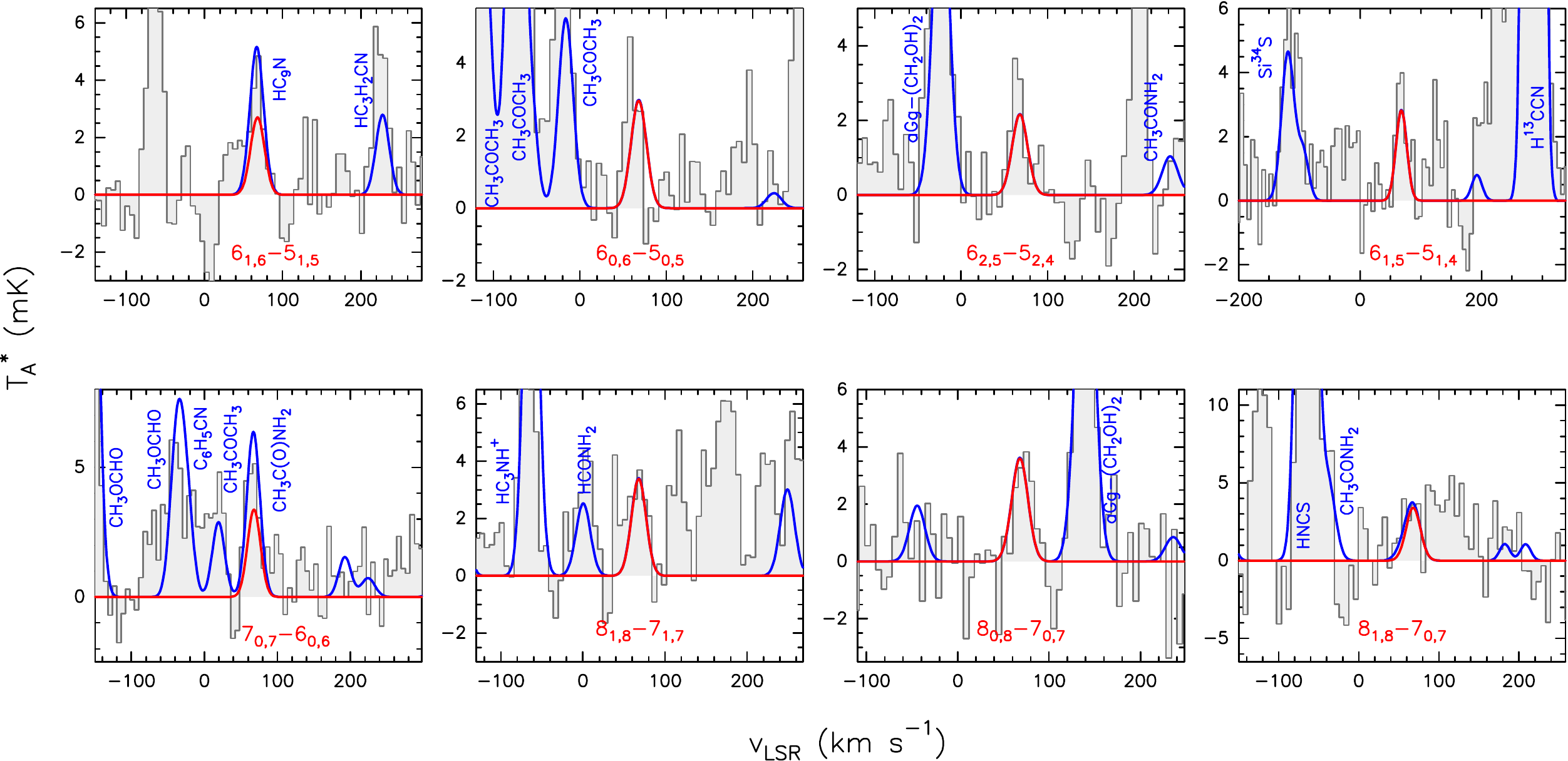}
\vspace{-2mm}
      \caption{Selected lines of EtNCO. The grey areas indicate the observed spectra smoothed up to 3~km~s$^{-1}$ for optimal line visualisation; while the red and blue lines represent the best LTE fit for the single EtNCO and all the other detected species, respectively. Blue labels are indicating the detected species within each spectral range while the quantum numbers of each EtNCO line are indicated in red. See the text and Table \ref{tab:c2h5nco_h2nco} for more details.
              }
         \label{fig:c2h5nco}
\end{figure*}

\section{Introduction}
More than 240 molecules\footnote{\url{https://cdms.astro.uni-koeln.de/classic/molecules}} have been detected in the interstellar medium (ISM).
Among them, only four are carrying the isocyanate functional group (-N=C=O). 
Namely: isocyanic acid (HNCO), one of the first detected  molecule in space \citep{snyder1972}; isocyanic radical (NCO) and N-protonated isocyanic acid (H$_2$NCO$^+$), only reported in the L483 dense core \citep{marcelino2018NCO}; and methyl isocyanate (CH$_3$NCO, hereafter MeNCO), reported in 
hot cores \citep[e.g. Sgr B2N, Orion KL, G10.47+0.03;][]{halfen2015,cernicharo2016,gorai2020G10}, hot corinos \citep[e.g. IRAS 16293-2422, Serpens SMM1;][]{ligterink2017,martindomenech2017,ligterink2021} and in the G+0.693-0.027 molecular cloud \citep[hereafter G+0.693;][]{zeng2018}.


Isocyanates are molecules with prebiotic interest, since they play a role in the 
synthesis of amino acids, polymerisation of peptides \citep{pascal2005}, and in the production of nucleotides \citep{choe2021} and nucleosides \citep{schneider2018noncanonical}.
Until now, MeNCO has been the most complex isocyanate detected in the ISM,
while the search of more complex species, such as ethyl isocyanate (C$_2$H$_5$NCO, hereafter EtNCO), only yielded upper limits \citep[][]{kolesnikova2018,colzi2021guapos}. 

In this Letter we describe the detection of EtNCO towards G+0.693. This cloud, located in the Galactic Centre within the Sgr B2 molecular complex, It contains a very rich chemical inventory
\citep[][]{requenatorres2008,zeng2018,rivilla2019,rivilla2020prebiotic,jimenez2020,rodriguezalmeida2021,rivilla2021ethanolamine}. 
G+0.693 is thought to be undergoing a cloud-cloud collision \citep[][]{zeng2020cloud}, which produces large-scale shocks that sputter dust grains, enhancing the gas-phase abundance of molecules by several orders of magnitude \citep[][]{requenatorres2006}. Since the abundances of HNCO are MeNCO in this cloud are relatively high 
\citep[>~10$^{-10}$;][]{zeng2018}, together with several detections of other ethyl derivatives, such as ethanol \citep[C$_2$H$_5$OH, hereafter EtOH;][]{requenatorres2006}, ethyl cyanide \citep[C$_2$H$_5$CN, hereafter EtCN;][]{zeng2018}, and recently ethyl mercaptan \citep[C$_2$H$_5$SH, hereafter EtSH;][]{rodriguezalmeida2021}, G+0.693 represents an excellent candidate for the search of EtNCO and other isocyanates.

\vspace{-3mm}
\section{Observations}

A high sensitivity spectral survey was carried out 
towards G+0.693.
We used both IRAM 30m (Granada, Spain) and Yebes 40m telescopes (Guadalajara, Spain). The observations were centred at $\alpha$(J2000.0)$\,$=$\,$17$^h$47$^m$22$^s$, and $\delta$(J2000.0)$\,$=$\,-\,$28$^{\circ}$21$'$27$''$. The position switching mode was used in all the observations with the off position located at an offset $\Delta\alpha$~=~$-885$'', $\Delta\delta$~=~$290$''.
For the IRAM 30m observations, the dual polarisation receiver EMIR was used connected to the fast Fourier transform spectrometers (FFTS), which provided a channel width of 200 kHz in the 
radio windows from 71.8 to 116.7 GHz and from 124.8 to 175.5 GHz. 
The observations with the Yebes 40m radiotelescope  (project number 20A008, PI Jim{\'e}nez-Serra)
used the Nanocosmos Q-band (7$\,$mm) HEMT receiver
\citep[][]{tercero2021yebes}. The receiver was connected to 16 FFTS providing a channel width of 38 kHz and a bandwidth of 18.5 GHz per lineal polarisation, covering the frequency range between 31.3 GHz and 50.6 GHz. See \cite{zeng2020cloud} and \cite{rivilla2021ethanolamine} for a more detailed description of the observations. 

\vspace{-3mm}
\section{Analysis and results}
\label{sec:analysis}

For the 
line identification of the molecular species, we have used the \textsc{madcuba} software\footnote{\textsc{madcuba} is a software developed by the Centro de Astrobiolog\'{\i}a (CSIC-INTA) located in Madrid (Spain). \url{https://cab.inta-csic.es/madcuba/index.html}.}. This is supported by the Spectral Line Identification and Modelling (SLIM) tool, which generates a synthetic spectra under the assumption of Local Thermodynamic Equilibrium (LTE) conditions and also considering line opacity effects 
(\citealt{martin2019}). 
The free parameters of the fits
are: the column density ($N$), the excitation temperature ($T_{\mathrm{ex}}$), the full width at half maximum ($FWHM$) and the local standard of rest velocity ($v_{\mathrm{LSR}}$). 
For the LTE analysis of each molecular species, we have also considered the emission from other molecules previously identified in our spectral survey that could potentially produce line contamination \citep{requenatorres2006,requenatorres2008,zeng2018,rivilla2019,jimenez2020,rivilla2020prebiotic,rivilla2021ethanolamine,rivilla2021HNCN,rodriguezalmeida2021}.

For each molecule analysed in this work, we have listed the details of the spectroscopy in Appendix \ref{appendixA:spectroscopicinfo}.



\vspace{-3mm}
\subsection{Detection of EtNCO}


Figure \ref{fig:c2h5nco} shows the brightest lines of EtNCO (details of the spectroscopy in Table \ref{tab:spectroscopicinfo}) detected at levels above 5$\sigma$ in integrated intensity towards G+0.693, which add up to a total of 8 transitions. Five of them appear unblended, while the other three are slightly blended with known (and identified) molecular species (see Table \ref{tab:c2h5nco_h2nco}). The contamination from unidentified lines is negligible for these transitions, except for the line at 33.966 GHz which shows a small excess with respect to the LTE fit. The remaining lines of EtNCO with line intensities predicted by the LTE fit at levels above 5$\sigma$ are strongly blended with emission of other species. It is worth mentioning that all clear transitions have been detected in the 7~mm radio window, which is less affected by line overlaps.
The LTE fit was carried out by fixing FWHM and v$_{\mathrm{LSR}}$ to 21~km~s$^{-1}$ and 68~km~s$^{-1}$, respectively, giving a $T_{\mathrm{ex}}\,$=$\,$10$\,\pm\,$2~K and $N$~=~(8.1$\,\pm\,$1.3)$\,\times\,$10$^{12}$~cm$^{-2}$,
To calculate the molecular abundance with respect to H$_2$ (all the abundances through the text are given with respect to H$_2$, unless otherwise mentioned), we have used $N_{\mathrm{H_2}}$~=~1.35$\times$10$^{23}$~cm$^{-2}$ derived by \citet{martin2008}. 
To estimate the uncertainties of the molecular abundance we have considered the error in the column density obtained with \textsc{madcuba} with an additional 15\% of uncertainty on account of calibration errors. We obtain a molecular abundance of (4.7$-$7.3)$\times$10$^{-11}$.
With the results obtained from EtNCO and the derived column density of MeNCO towards G+0.693 \citep{zeng2018}, a ratio MeNCO~/~EtNCO~=~8$\,\pm\,$1 is obtained.


\begin{table*}[h]
\setlength{\tabcolsep}{2pt}
\caption{\label{tab:c2h5nco_h2nco} Spectroscopic and information of the LTE fitting of the selected lines of EtNCO shown in Figure \ref{fig:c2h5nco}. The frequency, quantum numbers (QNs) the energy of the upper level (E$_{\mathrm{u}}$), logarithm of the Einstein coefficients ($\log \mathrm{A_{ul}}$), root mean square (rms) of the analysed spectral region, integrated intensity ($\int T_A^* d\nu$) and signal-to-noise ratio in integrated intensity (S/N) are given.}
\centering
\renewcommand{\arraystretch}{1.0}
\begin{tabular}{cccccccc}
\hline
Frequency & QNs
& E$_{\mathrm{u}}$ &$\log\mathrm{A_{ul}}$ & rms &$\int T_A^* d\nu$&S/N\tablefootmark{b} & blending \\
(GHz)&(J$''_{\mathrm{K_a'',K_c''}}$- J$'_{\mathrm{K_a',K_c'}}$)& (K) &(s$^{-1}$)&(mK)&($\mathrm{mK\,km\,s^{-1}}$)\tablefootmark{a}& & \\
\hline 
32.537109 & $6_{1,6}-5_{1,5}$ & 6.0  & -5.4198 & 1.5 & 61 & 7 & blended with HC$_9$N \\
33.545973 & $6_{0,6}-5_{0,5}$ & 5.7  & -5.3818 & 1.2 & 67 & 10 & clean  \\
33.966567 & $6_{2,5}-5_{2,4}$ & 7.9  & -5.4239 & 1.2 & 49 & 7 & clean   \\ 
35.293167 & $6_{1,5}-5_{1,4}$ & 6.5  & -5.3498 & 1.3 & 63 & 9  & clean \\
38.944488 & $7_{0,7}-6_{0,6}$ & 7.5 & -5.1881 & 1.2 & 76 & 11   &  blended with CH$_3$C(O)NH$_2$ \\
43.270670 & $8_{1,8}-7_{1,7}$ & 9.9  & -5.0473 & 1.6 & 76 & 8    & clean \\
44.274655 & $8_{0,8}-7_{0,7}$ & 9.7  & -5.0220 & 1.5 & 82 & 10   & clean \\
46.895390 & $8_{1,8}-7_{0,7}$ & 10.7 & -5.3581 & 2.7 & 77 & 5 &  blended with unidentified species \\
\hline
\end{tabular}
\vspace{-3mm}
\tablefoot{
\tablefoottext{a}{Integrated intensity derived from \textsc{madcuba} LTE fit, see Section~\ref{sec:analysis} for details.}
\tablefoottext{b}{S/N$\,$=$\mathrm{\Big(\,\int T_A^* d\nu\Big)\;/\;\Big[rms\,\bigg(\frac{\Delta v}{\mathrm{FWHM}}\bigg)^{0.5}\mathrm{FWHM}\Big]}$}
, where $\Delta \nu$ represents the spectral resolution of the data in velocity units.}
\end{table*}

\begin{figure*}[h]
   \centering
       \includegraphics[width=15cm]{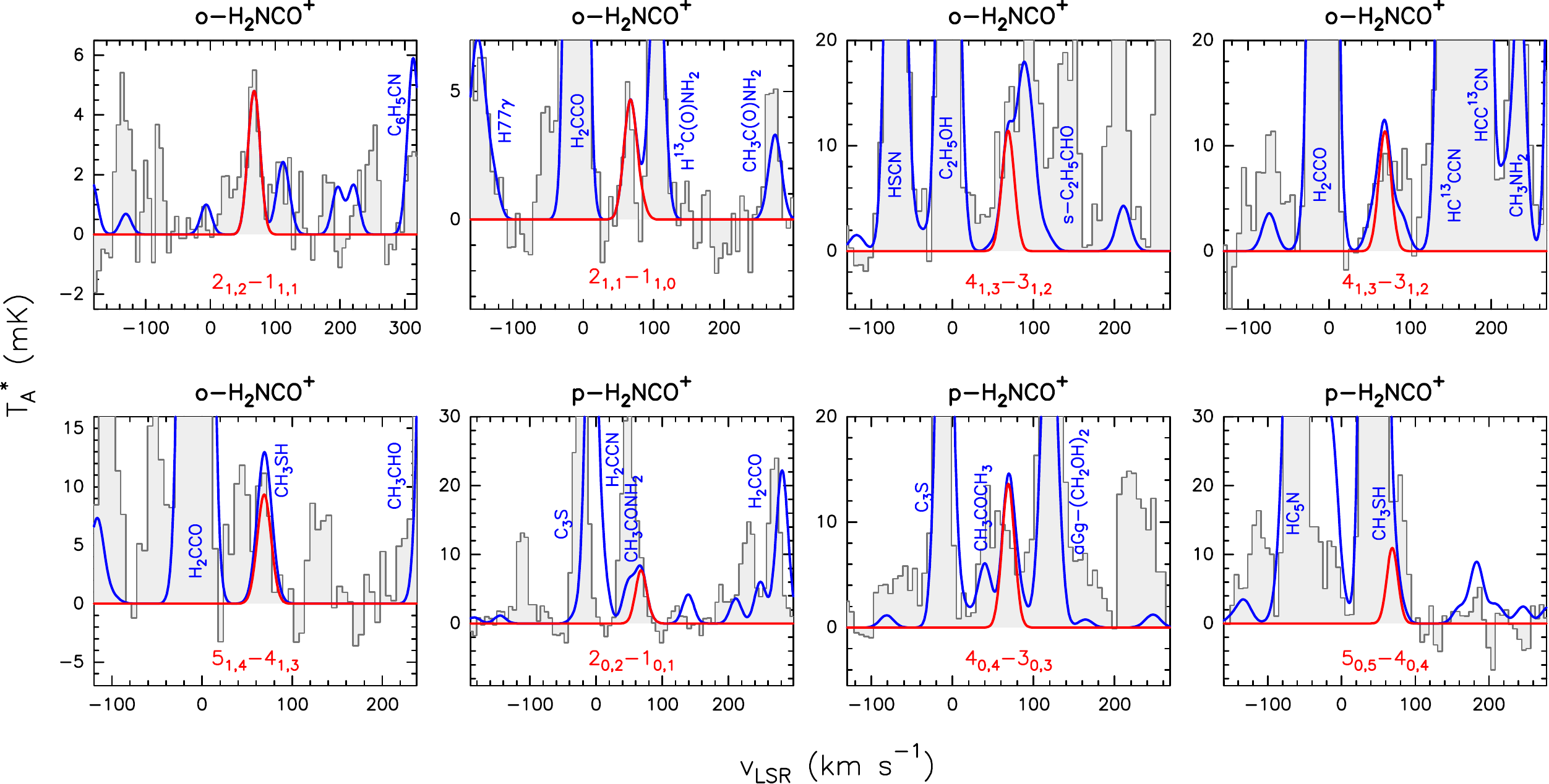}
 \vspace{-1.5mm}
      \caption{\label{fig:h2nco} Selected lines of H$_2$NCO$^+$. See Figure~\ref{fig:c2h5nco} for a description of the information given in the plot. In this case, the LTE fit was carried out separating ortho (o-H$_2$NCO$^+$) and para (p-H$_2$NCO$^+$) species (see the text for details) which are indicated above each panel. See Table \ref{tab:h2nco+} for a complete description of the quantum numbers.}
\end{figure*}

\vspace{-3mm}
\subsection{Detection of H$_2$NCO$^+$ and search for other isocyanates}

H$_2$NCO$^+$ (see Table \ref{tab:spectroscopicinfo} for the spectroscopic references) has been already reported towards L483 \citep{marcelino2018NCO} and tentatively detected towards Sgr B2 \citep{gupta2013}. 
Since the $T_{\mathrm{ex}}\,$ towards G+0.693 is low, we have separated ortho and para states for the analysis \citep[energy difference $\sim$15~K;][]{gupta2013}.

In Figure \ref{fig:h2nco} and Table \ref{tab:h2nco+} we show the H$_2$NCO$^+$ transitions detected towards G+0.693: 6 transitions of the ortho state (K$_{\mathrm{a}}$ odd, o-H$_2$NCO$^+$), 3 of which are unblended, and 3 transitions of the para state (K$_{\mathrm{a}}$ even, p-H$_2$NCO$^+$) which appear blended with other species. The LTE fits of the ortho and para states have been obtained separately.
For o-H$_2$NCO$^+$, we fixed $v_{\mathrm{LSR}}\,$ and FWHM at 69 and 18 km~s$^{-1}$, respectively, obtaining $N$~=~(1.1$\,\pm\,$0.2)$\,\times\,$10$^{12}$~cm$^{-2}$
and $T_{\mathrm{ex}}\,$=~7$\,\pm\,$1~K. For the para states, we have assumed the same $T_{\mathrm{ex}}$ and obtained $N$~=~(6.3$\,\pm\,$1.7)$\,\times\,$10$^{11}$~cm$^{-2}$. 
The total $N$ (ortho + para) gives (1.7$\,\pm\,$0.2)$\,\times\,$10$^{12}$~cm$^{-2}$ and an abundance of (1.1$-$1.5)$\,\times\,$10$^{-11}$ 
(Figure \ref{fig:all_isocyanates}).


We have also searched for other isocyanates, namely: isocyanate radical (NCO) $-$ previously found towards L483 \citep{marcelino2018NCO} $-$ vinyl isocyanate (C$_2$H$_3$NCO), cyanogen isocyanate (NCNCO) and ethynyl isocyanate (HCCNCO) with the references used to search for them listed in Table \ref{tab:spectroscopicinfo}.
Since none of them were detected, a 3$\,\sigma$ upper limit to their column densities have been measured using in each case the cleanest and brightest transition in the dataset (presented in the Appendix \ref{AppendixB:upperlimits}). The derived upper limits are $\,$<$\,$1.4$\,\times\,$10$^{13}\,$cm$^{-2}$ for NCO,
$\,<\,$1.2$\,\times\,$10$^{12}\,$cm$^{-2}$ for NCNCO,  $\,<\,$2.5$\,\times\,$10$^{12}$cm$^{-2}$ for trans-C$_2$H$_3$NCO, $\,<\,$9.3$\,\times\,$10$^{12}$cm$^{-2}$ for cis-C$_2$H$_3$NCO and <$\,$8.9$\,\times\,$10$^{12}$cm$^{-2}$ for HCCNCO (see Table \ref{tab:upperlimitslines}).

\vspace{-3mm}
\section{\label{sec:discussion} Discussion}


\subsection{\label{subsec:HNCOproduction} Isocyanate family in G+0.693}

\begin{figure}[htpb!]
    \centering
    \includegraphics[scale=0.75]{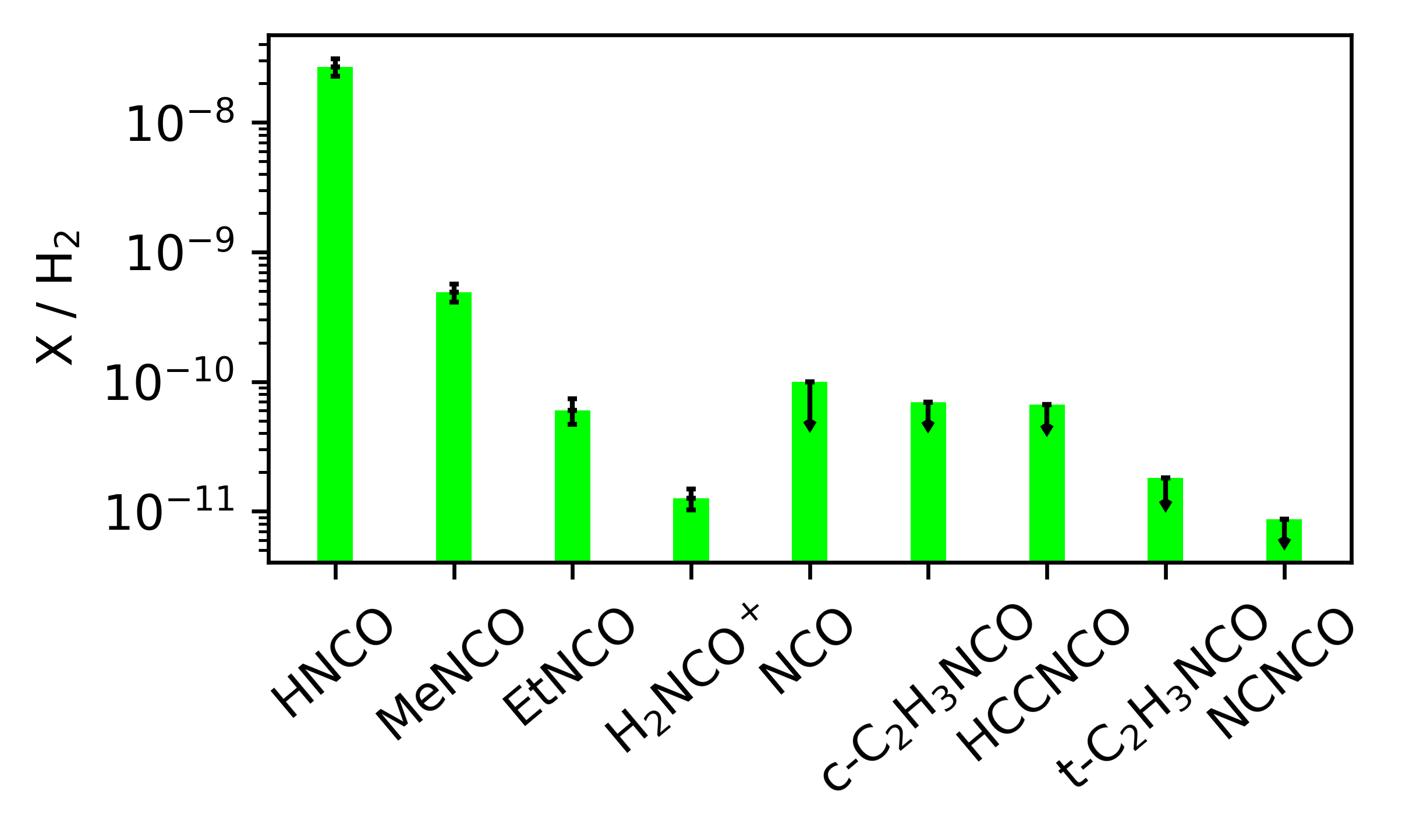}
    \vspace{-4mm}
    \caption{Molecular abundances with respect to H$_2$ of the isocyanates targeted towards G+0.693. 
    }
    \label{fig:all_isocyanates}
\end{figure}

We have plotted in Figure \ref{fig:all_isocyanates} the abundances of the isocyanates already detected in G+0.693 \citep[HNCO and MeNCO;][]{zeng2018} together with the ones reported in this work (detected: EtNCO and H$_2$NCO$^+$, and non-detected upper limits: NCO, cis/trans-C$_2$H$_3$NCO and NCNCO). 
Among the detected species, the molecular abundances decrease when increasing the chemical complexity (Figure \ref{fig:all_isocyanates}): HNCO $>$ MeNCO $>$ EtNCO.  
Contrary, the simplest species, the radical NCO, is not detected, with an upper limit that is a factor of 5 lower than the abundance of MeNCO. The high reactivity of this radical could explain its low gas-phase abundance. 


The ratio ethyl/vinyl (i.e. EtNCO~/~C$_2$H$_3$NCO) gives >$\,$3.7 and >$\,$1.5 for the cis and trans isomers of C$_2$H$_3$NCO, respectively. This is in contrast with the ethyl/vinyl ratio found towards G+0.693 in cyanides (C$_2$H$_5$CN/C$_2$H$_3$CN$\sim$0.5, \citealt{zeng2018}) and amines (C$_2$H$_5$NH$_2$~/~C$_2$H$_3$NH$_2\sim$0.7, Zeng et al. submitted).

The ratio between HNCO and its protonated species is H$_2$NCO$^+$~/~HNCO is $\sim$0.5$\times$10$^{-3}$, which is around five times lower than that found towards the cold core L483 of $\sim$2.5$\times$10$^{-3}$ (\citealt{marcelino2018NCO}). This might indicate that the H$_2$NCO$^+$~/~HNCO ratio decreases with the kinetic temperature of the source ($\sim$50$-$150 K in G+0.693 and $\sim$10 K in L483; \citealt{zeng2018} and \citealt{agundez2019}, respectively). A similar behaviour has been recently observed for HCNH$^+$~/~HCN  towards a sample of massive star-forming regions, where the ratio varies from 0.1$-$0.05 in starless (i.e. cold) sources \citep[in agreement with the value found in the low-mass cold core L1544 of $\sim$0.05 to (1$-$5)$\times$10$^{-3}$ in protostellar (i.e. hot) sources;][respectively]{quenard2017HC3NH+,fontani2021}. These authors proposed that the distinct values are due to different formation routes of HCNH$^+$ at low and hot temperatures.  Further observations of H$_2$NCO$^+$ towards new sources, and dedicated chemical models (see also Section \ref{subsec:NCOchemistry}) are needed to understand how the  H$_2$NCO$^+$~/~HNCO varies with the gas temperature.



\begin{figure*}[htb!]
    \centering
    \includegraphics[width=\textwidth]{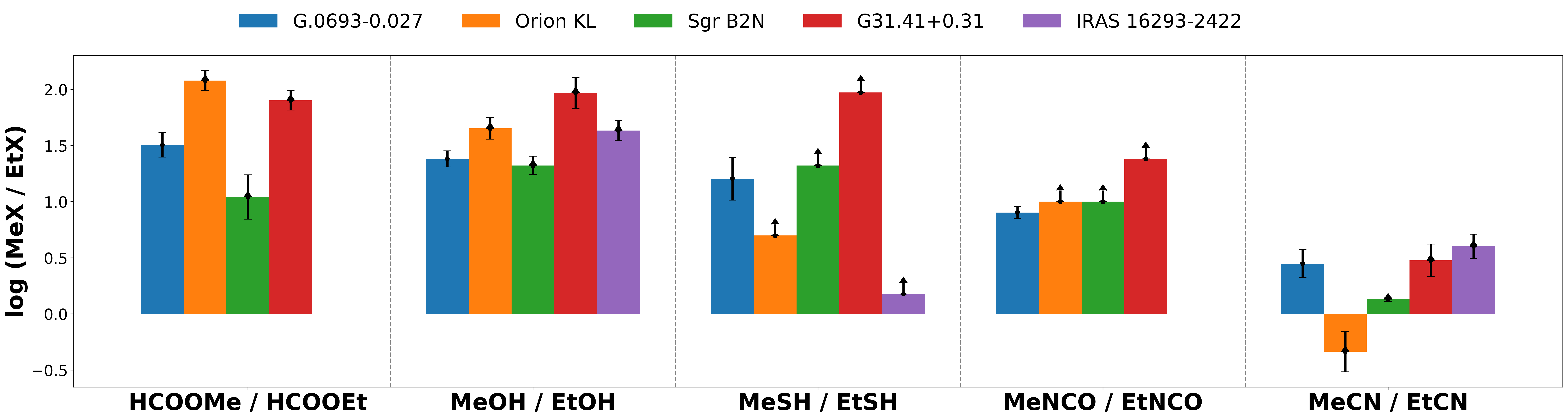}
    \vspace{-5mm}
    \caption{Comparison of MeX~/~EtX ratios among different regions of the ISM; where X~=~HCOO, OH, SH, NCO and CN (from right to left). The data has been taken from: \citet{zeng2018}, \citet{rodriguezalmeida2021} and this work for G+0.693-0.027; \citet{kolesnikova2014}, \citet{lopez2014}, \citet{tercero2015} and \citet{kolesnikova2018} for Orion KL; \citet{belloche2009} and \citet{kolesnikova2018} for Sgr B2N; \citet{mininni2020}, \citet{colzi2021guapos} and Mininni et al. (in prep) for G31.41+0.31; \citet{drozdovskaya2019} for IRAS 16293-2422.}
    \label{fig:me/etcomparison}
\end{figure*}

\vspace{-3mm}
\subsection{\label{subsec:Me/Et ratios} Methyl-to-ethyl ratios in the ISM}

The detection of EtNCO in the ISM, and the recent confirmation of EtSH (\citealt{rodriguezalmeida2021}), led us to study the methyl~/~ethyl (Me~/~Et) ratios in different molecular families in G+0.693 and other interstellar regions. In Figure \ref{fig:me/etcomparison} we have compared the Me~/~Et ratio in G+0.693 with those found in the massive hot cores G31.41+0.31, Orion KL, and Sgr B2N, and the low-mass hot corino IRAS 16293-2422. We have considered different chemical families: alcohols (-OH), nitriles (-CN), formiates (-HCOO), thiols (-SH) and isocyanates (-NCO).

Figure \ref{fig:me/etcomparison} shows that in G+0.693 the observed ratio MeNCO~/~EtNCO$\sim$8 is 2-4 times lower than the ratios for the formiates, alcohols and thiols (32, 24 and 16, respectively) and $\sim$4 times higher than the MeCN~/~EtCN ratio. 

Within each source, the Me~/~Et ratio of formiates, alcohols and thiols fall within the same range considering the uncertainties. However, the Me~/~Et ratio is clearly lower in the N-bearing species: isocyanates (for which only the G+0.693 value is available) and nitriles. This suggests a more efficient production of EtNCO and EtCN, compared to their methyl-counterparts, than others ethyl-derivatives.

\vspace{-3mm}
\subsection{\label{subsec:NCOchemistry} Formation of C$_2$H$_5$NCO and H$_2$NCO$^+$ and in the ISM}

The interstellar formation of the simplest 
isocyanates, HNCO and CH$_3$NCO, have been studied both theoretically (e.g. \citealt{quan2010,cernicharo2016,martindomenech2017,quenard2018,majumdar2018}) and experimentally \citep{ligterink2017,mate2018}. 
However, little is known about EtNCO and H$_2$NCO$^+$. 


Regarding EtNCO, there is no reaction proposed in UMIST \citep{mcelroy2013umist} or KIDA \citep[KInetic Database for Astrochemistry;][]{wakelam2012} chemical databases. \citet{sewilo2019} proposed the ion-molecule reaction in the gas phase C$_2$H$_5$OH$_2^+$ + HNCO $\rightarrow$ C$_2$H$_5$NCO + H$_2$O.
Despite the viability of the reaction, the abundance of a protonated species normally decreases by few orders of magnitude with respect to the non-protonated one, as we have seen e.g. with H$_2$NCO$^+$ and HNCO towards G+0.693 (H$_2$NCO$^+$/ HNCO$\sim$0.5$\times$10$^{-3}$).
Since the abundance of EtOH towards G+0.693 is $\sim\,$5$\,\times\,$10$^{-9}$ \citep{requenatorres2006}, the expected abundance of C$_2$H$_5$OH$_2^+$ would be of the order of 10$^{-12}$, below the derived abundance of EtNCO. Therefore, it seems unlikely that C$_2$H$_5$OH$_2^+$ would be a progenitor of EtNCO.
Another possible formation pathway could involve analogous routes to that proposed for MeNCO. As discussed by \citet{majumdar2018}, MeNCO can be efficiently formed  on the dust grains by the radical-radical reaction CH$_3$ + NCO $\rightarrow$ CH$_3$NCO, which has been also proven experimentally \citep[][]{ligterink2017}. Therefore, the analogous grain surface reaction C$_2$H$_5$ + NCO $\rightarrow$ C$_2$H$_5$NCO might play an important role in the production of this species, although further experimental and/or theoretical work is needed to confirm this hypothesis. 

Regarding H$_2$NCO$^+$, \citet{marcelino2018NCO} suggested the thermodynamically possible ion-molecule reaction NH$_3$ + HCO$^+\,\rightarrow\,$H$_2$NCO$^+$ + H$_2$. 
This process is exothermic but clearly competes with the -probably- faster acid-base channel, i.e., HCO$^+$ + NH$_3$ $\rightarrow$ CO + NH$_4^+$.
Another possible reaction, proposed in the UMIST database, involves the ion-molecule reaction HNCO$^+$ + H$_2$ $\rightarrow$ H$_2$NCO$^+$ + H.
HNCO$^+$ could be obtained by photoionization \citep[HNCO first ionization potential $\sim\,$11.60 eV;][]{holzmeier2015HNCO_PI} through secondary photons, since the cosmic-ray ionisation rate is high in the Galactic Centre \citep{goto2013cosmic}. A similar ion-molecule reaction was proposed by \citet{iglesias1977} with NCO$^+$ and H$_2$. However, none of these ions are detected in the ISM up to date.
Alternative routes involve the gas phase proton transfer with a more acidic compound: HNCO + HX$^+$ $\rightarrow$ H$_2$NCO$^+$ + X. The best candidate is H$_3^+$ due to its high abundance in the ISM \citep{oka2006} and considering that its gas basicity is lower than HNCO \citep[4.4 versus 7.8 eV, respectively;][]{hunter1998}. This reaction is also proposed in both KIDA and UMIST databases with a temperature-dependent reaction rate that decreases in a factor of $\sim\,$3 when the temperature goes from 10~K to 100~K; which could explain the differences seen in the H$_2$NCO$^+$~/~HNCO ratios towards L483 and G+0.693.



\vspace{-3mm}
\section{Conclusions}

In this Letter we report the detection of EtNCO (for the first time in the ISM) and H$_2$NCO$^+$ towards the molecular cloud G+0.693-0.027. We derived molecular abundances of (6$\pm$2)$\times$10$^{-11}$ and (1.3$\,\pm\,$0.5)$\times$10$^{-11}$, respectively. While the formation of EtNCO is more likely to proceed on the surface of dust grains, H$_2$NCO$^+$ is possibly formed via the proton exchange of of H$_3^+$ with HNCO, as suggested by the change in the abundance ratio H$_2$NCO$^+$~/~HNCO with temperature.

We have also studied the full inventory of the isocyanate family towards G+0.693. The relative abundance of HNCO:MeNCO:EtNCO is 1:1/55:1/447, which implies a decrease by a factor of 10 progressively going from HNCO to MeNCO and to EtNCO.

We have compared the Me~/~Et ratios among functional groups derived towards G+0.693 and several Galactic hot cores~/~corinos. Within each source, the values of HCOOMe~/~HCOOEt and MeOH~/~EtOH are similar, which is also followed by the MeSH~/~EtSH ratio towards G+0.693. However, the Me~/~Et ratio for the N-bearing compounds (isocyanates and nitriles) are lower, which might indicate a more efficient production of their associated ethyl-derivatives.

\begin{acknowledgements}
We are grateful to the IRAM 30m and Yebes 40m telescope staff for help during the different observing runs. IRAM is supported by the National Institute for Universe Sciences and Astronomy/National Center for Scientific Research (France), Max Planck Society for the Advancement of Science (Germany), and the National Geographic Institute (IGN) (Spain). The 40m radio telescope at Yebes Observatory is operated by the IGN, Ministerio de Transportes, Movilidad y Agenda Urbana. L.F.R.-A., V.M.R. and L.C.  acknowledge support from the Comunidad de Madrid through the Atracción de Talento Investigador Modalidad 1 (Doctores con experiencia) Grant (COOL:Cosmic Origins of Life; 2019-T1/TIC-15379). I.J.-S. and J.M.-P. have received partial support from the Spanish State Research Agency (AEI) project number PID2019-105552RB-C41. We also acknowledge support from the Spanish National Research Council (CSIC) through the i-Link project number LINKA20353. PdV and BT thank the support from the European Research Council through Synergy Grant ERC-2013-SyG, G.A. 610256 (NANOCOSMOS) and from the Spanish Ministerio de Ciencia e Innovación (MICIU) through project PID2019-107115GB-C21. BT also thanks the Spanish MICIU for funding support from grant PID2019-106235GB-I00.

\end{acknowledgements}

%
\vspace{-8mm}
\bibliographystyle{aa} 
\bibliography{bibl.bib} 
%
\begin{appendix} 

\section{\label{appendixA:spectroscopicinfo} Spectroscopic information of the isocyanates searched towards G+0.693}

In Table \ref{tab:spectroscopicinfo} we have introduced all the molecules searched in our spectral survey, each one with its pertinent references for the line list and dipole moments.

For EtNCO, the spectral predictions were done re-evaluating the ro-vibrational partition function in a similar way as with MeNCO by \citet{cernicharo2016}  accounting only for the vibrational states below 2.6 GHz. See the details in \citealt{colzi2021guapos} (Appendix B.2).

\begin{table*}[htpb!]
\setlength{\tabcolsep}{7pt}
\caption{\label{tab:spectroscopicinfo} Spectroscopic information and references for the molecules studied in this work.}
\centering
\renewcommand{\arraystretch}{1.1}
\begin{tabular}{ccccc}
\hline

Molecule & Catalogue & Tag
& Dipole moment & Line list   \\
\hline 
C$_2$H$_5$NCO & \textsc{madcuba}\tablefootmark{a} & $\ldots$ & \citet{sakaizumi1976} & \citet{heineking1994,kolesnikova2018} \\
H$_2$NCO$^+$ & CDMS & 044516 & \citet{lattanzi2012} & \citet{gupta2013} \\
NCO & CDMS & 042503 & \citet{saito1970} & \citet{kawaguchi1985} \\
cis-C$_2$H$_3$NCO & \textsc{madcuba}\tablefootmark{a} & $\ldots$ & \citet{bouchy1977geometrical} & \citet{kirby1978microwave}\\
trans-C$_2$H$_3$NCO & \textsc{madcuba}\tablefootmark{a} & $\ldots$ & \citet{bouchy1977geometrical,bouchy1979determination} & \citet{kirby1978microwave} \\
NCNCO & \textsc{madcuba}\tablefootmark{a} & $\ldots$ & \citet{hocking1976microwave} & \citet{hocking1976microwave} \\
HCCNCO & \textsc{madcuba}\tablefootmark{a} & $\ldots$ & Assumed\tablefootmark{b} & \citet{ross1992}  \\
\hline
\end{tabular}
\vspace{-3mm}
\tablefoot{\\
\tablefoottext{a}{The entry was imported into \textsc{madcuba} using the spectroscopic works listed in the table.}
\tablefoottext{b}{Since there is no value reported in the literature, the *.cat file was generated assuming $\mu_a=1.0\,$D}.}
\end{table*}

\section{\label{apendixA:H2NCO+} Observed transitions and line parameters derived from the LTE fit of H$_2$NCO$^+$}

In Table \ref{tab:h2nco+} we have listed the complete description of the lines of H$_2$NCO$^+$ plotted in Figure \ref{fig:c2h5nco}. The transition of o-H$_2$NCO$^+$ at 100.307 GHz ($5_{1,5}-4_{1,4}$, E$_{\mathrm{u}}\,$=$\,$13.5$\,$K) is not shown, and it has not been considered for the fit, because it is heavily blended with CH$_3$OCHO (line at 100.308 GHz; intensity $\sim\,$60$\,$mK vs. $\sim\,$7$\,$mK for H$_2$NCO$^+$).

Note that both ortho and para H$_2$NCO$^+$ CDMS entries incorporate the \ce{^14N} nuclear spin hyper-fine splitting, but in our data these lines are unresolved. Hence, for simplicity, the F quantum numbers have been omitted from Figure \ref{fig:h2nco} (see Table \ref{tab:h2nco+} for a complete description of the lines and quantum numbers).

\begin{table*}[htpb!]
\setlength{\tabcolsep}{8pt}
\caption{\label{tab:h2nco+} Spectroscopic and information of the LTE fitting of the selected lines of o-H$_2$NCO$^+$ and p-H$_2$NCO$^+$ shown in Figure \ref{fig:h2nco}. The frequency, the QNs, the energy of the upper level (E$_{\mathrm{u}}$), the logarithm of the intensity (at 300$\,$K), $\int T_A^* d\nu$ and S/N in integrated intensity of the lines are given.}
\centering
\renewcommand{\arraystretch}{1.0}
\begin{tabular}{cccccccc}
\hline

Molecule & Frequency & QNs\tablefootmark{a}
& E$_{\mathrm{u}}$ &$\log\,\mathrm{Intensity}$ &$\int T_A^* d\nu$\tablefootmark{b} &S/N\tablefootmark{b}\ \\
& (GHz)& & (K) & (nm$^2$ MHz) & ($\mathrm{K\,km\,s^{-1}}$) & &  \\
\hline 
o-H$_2$NCO$^+$ &  40.122511 & 2$_{1,2}$-1$_{1,1}$ F~=~1-~1  & 1.9  & -5.0506 & 0.194 & 23 \\
& 40.122797  & 2$_{1,2}$-1$_{1,1}$ F~=~1-~2               & 1.9  & -6.2267 &  $\ldots$     & $\ldots$     \\
& 40.123224 & 2$_{1,2}$-1$_{1,1}$ F~=~1~-~0              & 1.9  & -4.9256 &    $\ldots$     & $\ldots$     \\
& 40.123539 & 2$_{1,2}$-1$_{1,1}\,$F~=~3~-~2              & 1.9  & -4.3024 &   $\ldots$      & $\ldots$     \\
& 40.124590 & 2$_{1,2}$-1$_{1,1}$  F~=~2~-~1               & 1.9  & -4.5734 &   $\ldots$      &  $\ldots$    \\
& 40.124876  & 2$_{1,2}$-1$_{1,1}$ F~=~2~-~2               & 1.9  & -5.0505 &   $\ldots$      &   $\ldots$   \\
& 40.781503 & 2$_{1,1}$-1$_{1,0}$ F~=~1~-~0 & 2.0  & -4.9115 & 0.116 & 17 \\
& 40.782899 & 2$_{1,1}$-1$_{1,0}$ F~=~2~-~2               & 2.0  & -5.0364 &  $\ldots$       &  $\ldots$    \\
& 40.783205 & 2$_{1,1}$-1$_{1,0}$ F~=~3~-~2               & 2.0  & -4.2882 &   $\ldots$      &   $\ldots$   \\
& 40.784147 & 2$_{1,1}$-1$_{1,0}$ F~=~2~-~1               & 2.0  & -4.5593 &  $\ldots$       &  $\ldots$    \\
& 40.784622 & 2$_{1,1}$-1$_{1,0}$ F~=~1~-~1               & 2.0  & -5.0364 &   $\ldots$      &    $\ldots$  \\
& 80.246416 & 4$_{1,4}$-3$_{1,3}$ F~=~5~-~4 & 8.7  & -3.3697 & 0.211 & 31 \\
& 80.246532 & 4$_{1,4}$-3$_{1,3}$ F~=~3~-~2               & 8.7  & -3.6030  &   $\ldots$      &  $\ldots$    \\
& 80.246564  & 4$_{1,4}$-3$_{1,3}$ F~=~4~-~3               & 8.7  & -3.4849 &   $\ldots$      &  $\ldots$    \\
& 81.565492  & 4$_{1,3}$-3$_{1,2}$ F~=~3~-~2 & 8.8  & -3.5890  & 0.209 & 31 \\
& 81.565524  & 4$_{1,3}$-3$_{1,2}$ F~=~5~-~4  & 8.8  & -3.3557 &     $\ldots$    & $\ldots$     \\
& 81.565633  & 4$_{1,3}$-3$_{1,2}$ F~=~4~-~3               & 8.8  & -3.4709 &   $\ldots$      &  $\ldots$    \\
& 101.95584 & 5$_{1,4}$-4$_{1,3}$ F~=~4~-~3 & 13.7 & -3.0757 & 0.165 & 18 \\
& 101.95585 & 5$_{1,4}$-4$_{1,3}$ F~=~6~-~5 & 13.7 & -3.1660  &   $\ldots$      &  $\ldots$    \\
& 101.95591 & 5$_{1,4}$-4$_{1,3}$ F~=~5~-~4  & 13.7 & -4.5462 &   $\ldots$      & $\ldots$    \\
\hline
p-H$_2$NCO$^+$ &   40.453210& 2$_{0,2}$-1$_{0,1}$ F~=~1~-~0 & 2.9     & -4.8970 & 0.168 & 20      \\
&40.454746 & 2$_{0,2}$-1$_{0,1}$ F~=~3~-~2    & 2.9     & -4.1488 &   $\ldots$    &      $\ldots$            \\
& 40.454815 & 2$_{0,2}$-1$_{0,1}$ F~=~2~-~1    & 2.9     & -4.4199 &    $\ldots$  &     $\ldots$    \\
& 40.455616  & 2$_{0,2}$-1$_{0,1}$ F~=~1~-~0 & 2.9     & -4.7720 &   $\ldots$    & $\ldots$   \\
& 40.455778 & 2$_{0,2}$-1$_{0,1}$ F~=~2~-~2  & 2.9     & -4.8970 &   $\ldots$    &  $\ldots$  \\
& 80.906908 & 4$_{0,4}$-3$_{0,3}$ F~=~5~-~4  & 9.7     & -3.3131 & 0.261 &    \\
& 80.906934 & 4$_{0,4}$-3$_{0,3}$ F~=~4~-~3   & 9.7     & -3.4283 &   $\ldots$    & $\ldots$   \\
& 80.907001  & 4$_{0,4}$-3$_{0,3}$ F~=~3~-~2  & 9.7     & -3.5464 &   $\ldots$    &  $\ldots$  \\
& 101.13112  & 5$_{0,5}$-4$_{0,4}$ F~=~6~-~5 & 14.6    & -3.0434 & 0.213 &    \\
& 101.13114 & 5$_{0,5}$-4$_{0,4}$ F~=~5~-~4   & 14.6    & -3.1336 &    $\ldots$   & $\ldots$   \\
 & 101.13117  & 5$_{0,5}$-4$_{0,4}$ F~=~4~-~3  & 14.6    & -3.2251 &    $\ldots$   & $\ldots$   \\
 & 101.13223  & 5$_{0,5}$-4$_{0,4}$ F~=~5~-~5  & 14.6    & -4.5138 &  $\ldots$     &$\ldots$  \\
\hline
\end{tabular}
\vspace{-3mm}
\tablefoot{\\
\tablefoottext{a}{The CDMS entry accounts for the hyper-fine splitting because of the nuclear spin of \ce{^{14}N}. Hence the QNs are presented in this way: J$''_{\mathrm{K_a'',K_c''}}\,\rightarrow\,$J$'_{\mathrm{K_a',K_c'}}$ F~=~F$''\rightarrow\,$F$'$.}
\tablefoottext{b}{In our data, the hyper-fine splitting is unresolved. Thus, the S/N is calculated by the sum of the individual F$''\rightarrow\,$F$'$ components.}.}
\end{table*}

\section{\label{AppendixB:upperlimits} Selected lines for computing the upper limits}

In Table \ref{tab:upperlimitslines} we have listed the brightest and cleanest lines for the column density upper limit determination for the non-detected isocyanates, namely: NCO, cis/trans-C$_2$H$_3$NCO and NCNCO.

\begin{table*}[htpb!]
\setlength{\tabcolsep}{8pt}
\caption{\label{tab:upperlimitslines} Lines employed to compute the upper limits to the column density of the isocyanates that have not been detected towards G+0.693.}
\centering
\renewcommand{\arraystretch}{1.0}
\begin{tabular}{cccccc} 
\hline
Molecule & Frequency & QNs
& E$_{\mathrm{u}}$ & $\int T_A^* d\nu$ & N\ \\
& (GHz)&  & (K) & ($\mathrm{mK\,km\,s^{-1}}$) & (cm$^{-2}$)   \\
\hline 
NCO (\ce{^2}$\Pi_{3/2}$) & 81.404300 & 7/2$\,-\,$5/2 F~=~9/2$\,-\,$7/2 e & 6.7 & 25 & $<\,$1.4$\,\times\,$10$^{13}$ \\
 & 81.404813 & 7/2$\,-\,$5/2 F~=~9/2$\,-\,$7/2 f & 6.7 & 25 & $\ldots$ \\
c-C$_2$H$_3$NCO & 46.281730 & 8$_{2,7}$-7$_{2,6}$ & 13.1 & 46 & $<\,$9.3$\,\times\,$10$^{12}$ \\
t-C$_2$H$_3$NCO & 33.483408 & 7$_{0,7}$-6$_{0,6}$ & 6.4 & 21 & $<\,$2.5$\,\times\,$10$^{12}$ \\
NCNCO & 42.364035 & 8$_{0,8}$-7$_{0,7}$ & 9.2 & 19 & $<\,$1.2$\,\times\,$10$^{12}$ \\
HCCNCO & 35.291156 & 7$_{0,7}$-6$_{0,6}$ & 7.1  & 28 & <$\,$8.9$\,\times\,$10$^{12}$ \\
\hline
\end{tabular}
\end{table*}

\section{\label{appendix:isocyanate_family} Molecular abundances of the isocyanates searched towards G+0.693}

In Table \ref{tab:all_isocyanates} we list the values of the molecular abundances of the isocyanates studied in this work with respect to H$_2$, which are plotted in Figure \ref{fig:all_isocyanates}. Additionally, we have also included the values of the column density together with the errors derived from the LTE fit with \textsc{madcuba}. The uncertainties of the molecular abundances are considering the errors derived from the column density and an additional 15\% that consider the calibration errors. 

\begin{table}[h]
    \centering
    \caption{\label{tab:all_isocyanates} Molecular abundance ratios of the isocyanates searched towards G+0.693.}
    \renewcommand{\arraystretch}{1.0}
    \begin{tabular}{ccc}
     \hline 
         Molecule & $N_{\mathrm{X}}\,\pm\,\Delta N_{\mathrm{X}}$ & $N_{\mathrm{x}}$ / $N_{\mathrm{H_2}}$ \\
         & ($\times\,$10$^{13}\,$cm$^{-2}$) & \\
         \hline
         NCO & $<\,$1.4  & <$\,$1.0$\,\times\,$10$^{-10}$ \\
         
         HNCO\tablefootmark{a} & (3.6$\,\pm\,$0.1)$\,\times\,$10$^2$ &
         (2.3$-$3.1)$\,\times\,$10$^{-8}$ \\
         
         o-H$_2$NCO$^+$ & 1.1$\,\pm\,$0.2 & 
         (5.8$-$9.8)$\,\times\,$10$^{-12}$ \\
         
         p-H$_2$NCO$^+$ & (6.3$\,\pm\,$1.7)$\,\times\,$10$^{-2}$  &
         (3.3$-$6.1)$\,\times\,$10$^{-12}$  \\
         
        (o+p)-H$_2$NCO$^+$ & (1.7$\,\pm\,$0.2)$\,\times\,$10$^{-1}$  & (1.1$-$1.5)$\,\times\,$10$^{-11}$  \\
        
         NCNCO & $<\,$0.12  & <$\,$8.7$\,\times\,$10$^{-12}$  \\ 
         
         CH$_3$NCO\tablefootmark{a} & 6.6$\,\pm\,$0.4 & (4.1$-$5.7)$\,\times\,$10$^{-10}$  \\
         
         C$_2$H$_5$NCO & 8.1$\,\pm\,$1.3 & (4.7$\,\pm\,$7.3)$\,\times\,$10$^{-11}$  \\
         
         trans-C$_2$H$_3$NCO & $<\,$0.25 & <$\,$1.8$\,\times\,$10$^{-11}$   \\
         
         cis-C$_2$H$_3$NCO & $<\,$0.93  & <$\,$6.9$\,\times\,$10$^{-11}$  \\
         
         HCCNCO & <$\,$0.89  & <$\,$6.6$\,\times\,$10$^{-11}$  \\
       
\hline
    \end{tabular}
\vspace{-3mm}
\tablefoot{
\tablefoottext{a}{Data taken from \citet{zeng2018}.}}
\end{table}

\end{appendix}






   
  



\end{document}